\documentclass[twocolumn,showpacs,preprintnumbers,amsmath,amssymb,prc]{revtex4}

\usepackage{graphicx}
\usepackage{dcolumn}
\usepackage{bm}

\usepackage{mathptmx, courier, pifont}
\usepackage[scaled=0.92]{helvet}
\usepackage[T1]{fontenc}
\usepackage{textcomp}

\newcommand{\quarterthin}{\kern 0.0417em}

\begin{document}

\title{
Competition in rotation-alignment between high-$j$ neutrons and
protons in transfermium nuclei }

\author{
Falih Al-Khudair$^{1,2,3}$, Gui-Lu Long$^{1,2}$, Yang
Sun$^{4,5}$\footnote{corresponding author: sunyang@sjtu.edu.cn} }

\affiliation{
$^{1}$Department of Physics, Tsinghua University, Beijing 100084,
People's Republic of China \\
$^{2}$Center of Nuclear Theory, Lanzhou National Laboratory of
Heavy Ion Accelerator, Lanzhou 730000, People's Republic of China \\
$^{3}$Department of Physics, College of Education, Basrah
University, Basrah, Iraq \\
$^{4}$Department of Physics, Shanghai Jiao Tong University, Shanghai
200240, People's Republic of China \\
$^{5}$Institute of Modern Physics, Chinese Academy of Sciences,
Lanzhou 730000, People's Republic of China }

\begin{abstract}
The study of rotation-alignment of quasiparticles probes sensitively
the properties of high-$j$ intruder orbits. The distribution of very
high-$j$ orbits, which are consequences of the fundamental
spin-orbit interaction, links with the important question of
single-particle levels in superheavy nuclei. With the deformed
single-particle states generated by the standard Nilsson potential,
we perform Projected-Shell-Model calculations for transfermium
nuclei where detailed spectroscopy experiment is currently possible.
Specifically, we study the systematical behavior of
rotation-alignment and associated band-crossing phenomenon in Cf,
Fm, and No isotopes. Neutrons and protons from the high-$j$ orbits
are found to compete strongly in rotation-alignment, which gives
rise to testable effects. Observation of these effects will provide
direct information on the single-particle states in the heaviest
nuclear mass region.
\end{abstract}

\date{\today}
\maketitle

\section{Introduction}

The occurrence of superheavy elements (SHE) is attributed to the
nuclear shell effect because the macroscopic liquid-drop model would
predict that such heavy elements can not exist due to large Coulomb
repulsive force. The distribution of single-particle (SP) states as
a consequence of the shell effect has thus become the discussion
focus in the SHE problem. One important question has been where are
the next magic numbers in the superheavy mass region beyond the
known magic number 126 for neutrons and 82 for protons. The precise
location of the new magic numbers depends sensitively on the SP
structure. Theoretically, stability is predicted for the nuclei
close to the spherical shells with $N = 184$ and $Z = 114$ (also $Z
= 120$ or 126, depending on theoretical models employed), which
suggests the existence of "island of stability" \cite{Stoyer06} next
to the well-known doubly-magic nucleus $^{208}$Pb.

Exploring the island is the current goal in nuclear science. In the
past few years, researchers have made significant progress in
synthesis of new elements (for review, see Refs.
\cite{Hofmann00,Armbruster00,Hofmann01,Oganessian07}). Presently,
little is known about their structure. The heaviest nuclei for which
detailed spectroscopy measurement can be performed lie in the
transfermium mass region, as for instance, the Californium, Fermium,
and Nobelium isotopes \cite{Leino04,Rodi04,Green07,HG08}. These
nuclei, typically with $N\approx 100$ and $Z\approx 150-160$, are
not really SHE. However, they are at the gateway to the SHE region,
and furthermore, are well deformed. As one can clearly see form the
deformed SP spectra \cite{Chas77} that with deformation, the Fermi
surfaces of these nuclei are surrounded by some orbitals originating
from the subshells near the anticipated new magic numbers. Thus, the
study of these deformed transfermium nuclei may provide an indirect
way to access the SP states of the closed spherical shells, which
are of direct relevance to the location of the predicted island.

In-beam measurements for the transfermium region have been performed
for yrast $\gamma$-ray spectroscopy of even-even nuclei (for
example, $^{250}$Fm \cite{Fm250}, $^{252}$No \cite{No252},
$^{254}$No \cite{No254}). The data reveal that these nuclei are well
deformed. At low-spins near the ground state, they all exhibit very
similar collective behavior with regular rotational level sequence.
This tells us that near the ground state, these nuclei behave like a
heavy, rigid rotor. They show a strong collectivity, diluting any
individual role of single-particles. Therefore, not much information
can be extracted from these low-spin rotor states.

More useful information may be obtained through the study of
high-spin states with quasiparticle excitations. In fact, some
non-yrast and isomeric states have been observed (see, for example,
Refs.
\cite{Hofmann,Butler02,Eeck05,Rodi06,Tandel06,Sulignano07,Katori08,Greenlees08},
and for the most recent review, see Ref. \cite{HG08}). The yielded
data contain useful information on excited levels and configurations
of multi-quasiparticle states in this mass region, and moreover,
they test strictly current nuclear models that have been used for
prediction. There are several types of quasiparticle excitation. One
possibility is the study of $K$-isomers through the isomer
spectroscopy measurement \cite{Jones02}. The isomer study has become
an important branch of nuclear structure research \cite{AS05}. The
suggestion of Xu {\it et al.} \cite{Xu04} has made the study of
isomeric states in SHE more interesting. These authors suggested
that the occurrence of isomeric states in SHE can enhance the
stability because the multi-quasiparticle excitations decrease the
probability for both nuclear fission and $\alpha$-decay. In the
present paper, we concentrate on another possibility of
quasiparticle excitations; namely, we discuss rotation-alignment at
high spins along the yrast line.

On the theoretical side, the early study by Munitian {\it et al.} on
rotational structure in very heavy nuclei employed cranking
approximation based on a macroscopic-microscopic approach
\cite{Munt99}. The first microscopic calculation in the framework of
the Hartree-Fock-Bogoliubov (HFB) approximation was carried out by
Egido and Robledo in Ref. \cite{ER00}, in which properties of the
ground-state rotational band in $^{254}$No were discussed in detail.
Subsequent studies include the cranked HFB calculations with the
Skyrme force \cite{Dug01,Ben03}, cranked HFB calculations with the
Gogny force \cite{Gogny06}, cranked relativistic Hartree-Bogoliubov
method \cite{Afa03}, and very recently, cranked shell model with
particle-number-conserving treatment \cite{He08}. In all these
papers, cranking approximation was adopted to describe rotation and
discussions were carried out in the intrinsic frame. Alternatively,
Hess and Misicu \cite{Hess03} used the pseudo-symplectic description
for low-spin collective motions in superheavy nuclei.

The present work is based on the Projected Shell Model (PSM)
\cite{PSM}, and aims at understanding the role of high-$j$ intruder
orbits in the high-spin states of transfermium nuclei. It has been
shown that the PSM describes efficiently the high-spin phenomena
such as band-crossing, rotation-alignment, and band-bending in
moment of inertia in well deformed nuclei. The present study
systematically covers the isotopic chains of Californium, Fermium,
and Nobelium, for states up to angular momentum 30$\hbar$, well
beyond the first band-crossing. Our method is different from the
cranking approximations in that the PSM works in the laboratory
frame in terms of the configuration mixing, and the observables such
as electromagnetic transition rates can be unambiguously computed in
the PSM framework. The PSM is thus free from the well-known problem
occurring at the band-crossing region caused by the cranking
approximation \cite{Hamamoto76}. This is important for a theoretical
treatment on band-crossings because the current experiment on
transfermium nuclei is approaching the high-spin regions across the
first band-crossing.

The paper is organized as follows. In Sec. II, we outline the
theoretical mehtod. Systematic analysis of the rotational structure
along the yrast line for even-even transfermium nuclei is carried
out in Sec. III. Discussions about the present results and their
implications are given in Sec. IV. Finally, a summary is given in
Sec. V.

\section{Outline of the Projected Shell Model}

A successful description of deformed nuclei can be traced back to
the introduction of the Nilsson model \cite{Nilsson55}. In the
Nilsson model, nuclear states are described by considering nucleons
moving in a deformed potential. Deformed states are defined in the
body-fixed frame of reference in which the rotational symmetry is
broken. Although physics may be discussed in such an intrinsic
frame, the broken rotational symmetry should in principle be
recovered giving the fact that angular momentum is probably the most
important quantum number in nuclear physics. Restoration of
rotational symmetry is the first step of going beyond mean field,
which can be done by using the angular-momentum-projection method
\cite{RS80}. The next step follows closely the basic strategy of the
conventional shell model: The projected states are then regarded as
a new basis in which one builds many-body wave functions in the
laboratory frame, and a two-body shell model Hamiltonian is
diagonalized in the projected basis. This means that one can use the
deformed Nilsson SP states as an effective basis. Thus, the
difference between the conventional shell model and the approach
with angular-momentum-projection is that one employs a spherical
basis to construct shell model basis in the former and a (projected)
deformed basis in the latter.

The above idea of performing shell model calculations is practiced
by the Projected Shell Model \cite{PSM}, which has been proven
successful in the description of heavy, deformed nuclei. The
calculation procedure is as follows. The PSM first constructs its
shell-model basis by using the deformed Nilsson SP states (with the
Nilsson parameters given in Ref. \cite{BR85}) at a quadrupole
deformation $\varepsilon_2$. Pairing correlations are incorporated
into the Nilsson states by a BCS calculation. The consequence of the
Nilsson-BCS calculations defines a set of quasiparticle (qp) states
associated with the qp vacuum $\left|\phi(\varepsilon_2)\right>
\equiv \left|0\right>$. One then considers the following multi-qp
configurations for even-even nuclei
\begin{equation}
\left|\Phi_\kappa\right> = \left\{\left|0 \right>, \
\alpha^\dagger_{n_i} \alpha^\dagger_{n_j} \left|0 \right>,\
\alpha^\dagger_{p_i} \alpha^\dagger_{p_j} \left|0 \right>,\
\alpha^\dagger_{n_i} \alpha^\dagger_{n_j} \alpha^\dagger_{p_i}
\alpha^\dagger_{p_j} \left|0 \right> \right\} , \label{baset}
\end{equation}
where $\alpha^\dagger$ is the qp creation operator and the index $n$
($p$) denotes neutron (proton) Nilsson quantum numbers in orbitals.
The angular-momentum-projected multi-qp states serve as the building
blocks of our shell model basis, and the trail wave functions can be
written as a superposition of them:
\begin{equation}
\left|\Psi^I_M\right> = \sum_\kappa f^I_\kappa \hat
P^I_{MK_\kappa} \left|\Phi_\kappa\right> , \label{wavef}
\end{equation}
where $\hat P^I_{MK}$ is the angular-momentum projection operator
\cite{RS80}
\begin{equation}
\hat P^I_{MK} = {2I+1 \over 8\pi^2} \int d\Omega\,
D^{I}_{MK}(\Omega)\, \hat R(\Omega),
\end{equation}
and $\kappa$ labels the basis states. It is the angular momentum
projection \cite{PSM} that transforms the wave functions from the
intrinsic frame to the laboratory frame. Finally a two-body
Hamiltonian is diagonalized in the projected states (\ref{wavef})
and the diagonalization determines $f^I_\kappa$. For details of how
to perform a projection calculation, we refer to the PSM review
article \cite{PSM}.

\begin{table*}
\caption{Input deformation parameters ($\epsilon_2$) used in the
calculation.} \label{tab:1}
\begin{tabular}{c|cccccccccccc}
\hline\noalign{\smallskip}
Nucleus & $^{246}$Cf & $^{248}$Cf & $^{250}$Cf & $^{252}$Cf &
$^{250}$Fm & $^{252}$Fm & $^{254}$Fm & $^{256}$Fm & $^{252}$No &
$^{254}$No & $^{256}$No
& $^{258}$No \\
\hline\noalign{\smallskip}

$\epsilon_2$ & 0.255 & 0.260 & 0.265 & 0.245 &
               0.240 & 0.255 & 0.250 & 0.222 &
               0.224 & 0.260 & 0.250 & 0.230 \\

\hline\noalign{\smallskip}
\end{tabular}
\end{table*}

We use the pairing plus quadrupole-quadrupole ($QQ$) Hamiltonian
with inclusion of both the monopole- and quadrupole-pairing terms
\begin{equation}
\hat H = \hat H_0 - {1 \over 2} \chi \sum_\mu \hat Q^\dagger_\mu
\hat Q^{}_\mu - G_M \hat P^\dagger \hat P - G_Q \sum_\mu \hat
P^\dagger_\mu\hat P^{}_\mu . \label{hamham}
\end{equation}
In Eq. (\ref{hamham}), $\hat H_0$ is the spherical SP Hamiltonian.
The $QQ$-force strength $\chi$ is determined in such a way that it
holds a self-consistent relation \cite{PSM} with the quadrupole
deformation $\epsilon_2$ (given in Table I below). The
monopole-pairing strength $G_M$ is of the form
\begin{equation}
G_M = {{21.24 \mp 13.86{{N-Z}\over A}}\over A},
\end{equation}
with ``$-$" for neutrons and ``$+$" for protons. The
quadrupole-pairing strength $G_Q$ is assumed to be proportional to
$G_M$, with the proportionality constant 0.12. We note that $G_Q$ is
an adjustable parameter of the PSM \cite{PSM}, and the
quadrupole-pairing force has an effect of shifting the position of
rotation-alignment \cite{Sun95}. At present, no definite data can
say about the rotation-alignment in this mass region, and there is
no motivation for us to use $G_Q$'s other than the present one that
reasonably described known yrast states and $\gamma$-vabrational
states \cite{Sun08}. As the valence SP space, we include three major
shells, $N=5,6,7$ (4,5,6), for neutrons (protons). The deformed
Nilsson SP states are generated with deformation parameters
$\epsilon_2$ listed in Table I, which are either obtained from
experimental data, if available, or from mean-field calculations. We
note that deformation in nuclei is a model-dependent concept. Our
deformations are input parameters for the deformed basis, and in
principle, it is not required that the numbers in Table I are
exactly the same as deformation parameters used in other models.
Nevertheless, it turns out that our employed deformation parameters
Ref. \cite{Sun08} for this mass region are very close to those
calculated in Refs. \cite{def1,def2}, and follow the same variation
trend along an isotopic chain as predicted by other models (for
example, the most deformed isotope has the neutron number 152 and a
decreasing trend for heavier isotopes is expected).

Once the wave functions of Eq. (\ref{wavef}) are obtained, one can
use them to directly calculate electromagnetic transition
probabilities \cite{Sun94}. The B(E2) value that measures the
electric quadrupole transition rate from an initial state $I$ to a
final state $I-2$ is given by
\begin{equation}
B(E2, I\rightarrow I-2) = \frac {1}{2I + 1} \left| \right<
\Psi^{I-2} || \hat Q_2 ||\Psi^I \left> \right|^2, \label{BE2}
\end{equation}
where wave functions $\left|\Psi^I\right>$ are those in Eq.
(\ref{wavef}). The effective charges used in our calculation are the
standard ones: $e^\pi=1.5e$ and $e^\nu=0.5e$, which are fixed for
all nuclei studied in this paper. Thus any variations in calculated
B(E2)'s are subject to the structure change in wave functions.

The gyromagnetic factor ($g$ factor) is the quantity most sensitive
to the SP components in wave functions as well as to their interplay
with collective degrees of freedom. Because of the intrinsically
opposite signs of the neutron and proton $g_s$, a study of $g$
factors enables the determination of the microscopic structure for
underlying states. For example, variation of $g$ factors often is a
clear indicator for a SP component that strongly influences the
total wave function. In the PSM, $g$ factors can be directly
computed by
\begin{equation}
g(I) = \frac {\mu(I)}{\mu_N I} = \frac {1}{\mu_N I} \left[ \mu_\pi
(I) + \mu_\nu (I) \right],
\label{gfactor}
\end{equation}
with $\mu_\tau (I)$ being the magnetic moment of a state
$\left|\Psi^I\right>$, expressed as
\begin{eqnarray}
\mu_\tau (I) &=& \left< \Psi^I_{I} | \hat
\mu^\tau_z | \Psi^I_{I} \right> \nonumber\\
&=& {I\over{\sqrt{I(I+1)}}} \left<
\Psi^{I} || \hat \mu^\tau || \Psi^{I} \right> \nonumber\\
&=& \frac{I}{\sqrt{I(I+1)}} \left[
     g^{\tau}_l \langle \Psi^I || \hat j^\tau ||\Psi^I
     \rangle + (g^{\tau}_s - g^{\tau}_l)
     \langle \Psi^I || \hat s^\tau || \Psi^I \rangle \right]
     , \nonumber
\label{moment}
\end{eqnarray}
where $\tau = \pi$ and $\nu$ for protons and neutrons, respectively.
The following standard values for $g_l$ and $g_s$ appearing in the
above quation are taken:
\begin{eqnarray*}
g_l^\pi &=& 1, ~~~~ g_s^\pi = 5.586 \times 0.75 ,\\
g_l^\nu &=& 0, ~~~~ g_s^\nu = -3.826 \times 0.75 .
\end{eqnarray*}
$g_s^\pi$ and $g_s^\nu$ are damped by a usual 0.75 factor from the
free-nucleon values to account for the core-polarization and
meson-exchange current corrections \cite{Castel90}. These same
values are used for all $g$ factor calculations in the present
paper, as in the previous PSM calculations, without any adjustment.

\section{Analysis of rotational structure along the yrast line}

The nucleon response to the collective rotation is generally
understood as follows. Near the non-rotating ground state of a
nucleus, like-nucleons are paired and interact coherently to form a
superfluid system. When a nucleus is rotating, the Coriolis
anti-pairing force acts on the pairs, and the force increases with
rotation. At a certain critical angular momentum one would expect
that the pairs are all broken, and a phase transition from the
superfluid to the normal-state phase would be observed
\cite{Mottelson60}. However, nucleons move in orbits. Nucleons in
different orbitals feel the Coriolis force very differently.
Detailed analysis has shown that the Coriolis force is proportional
to the size of the SP angular momentum $j$ of a nucleon under
consideration. Nucleons in the vicinity of the Fermi surface usually
occupy several orbitals with different $j$-values, and one expects
those nucleons with the highest $j$-value to break first and align
their rotation along the rotational axis \cite{Stephens72}. This is
the real situation happening in a rotating nuclear system, known as
the Stephens-Simon effect. It suggests that near the yrast line,
rotation-alignment of particular high-$j$ particles rather than a
collapse of entire pairing correlation is the dominant mode. These
high-$j$ orbitals are usually the intruder states that have an
opposite parity to their neighboring ones. Rotation-alignment lowers
the energy of the high-$j$ configurations, and at a certain angular
momentum, the states with aligning nucleons becomes so low
energetically that the qp aligning bands can cross the ground state
band (g-band), constituting a situation of so-called band-crossing.

\begin{figure*}
\includegraphics[width=15cm]{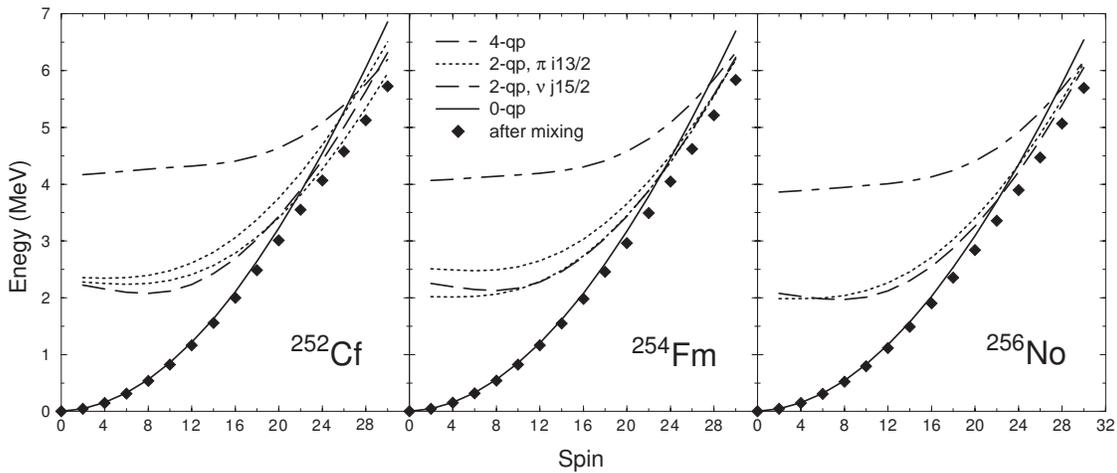}
\caption{Band diagrams for $^{252}$Cf, $^{254}$Fm, and $^{256}$No.
Several important configurations are shown: 0-qp (solid curves)
starting from the origin, neutron 2-qp (dashed curves) and proton
2-qp (dotted curves) starting from 2 -- 2.5 MeV, and 4-qp
(dotted-dashed curves) starting from about 4 MeV. Filled diamonds
denote the yrast states obtained after configuration mixing. }
\label{fig1}
\end{figure*}

Rotation-alignment is a well-known phenomenon in nuclear high-spin
physics, which is expected to occur also in rotating transfermium
nuclei that the present work studies. The characteristic feature of
rotation-alignment in transfermium nuclei is that the aligning pairs
come mainly from the two high-$j$ intruder orbitals: proton pairs
from the $i_{13/2}$ orbital and neutron pairs from the $j_{15/2}$
orbital. As we shall discuss below, the presence of these two
high-$j$ orbitals near the Fermi level and their response to
rotation can lead to an interesting competing picture at the band
crossing region, which may lead to observations.

\subsection{Band diagram}

A band diagram is a useful tool introduced in Ref. \cite{HS91} to
analyze the numerical results of the PSM. Each
angular-momentum-projected state in Eq. (\ref{wavef}) represents a
rotational band. The first one, $\hat P^I_{M0} \left| 0\right>$,
represents the g-band in which all the particles are paired. The
remaining states represent bands built upon multi-quasiparticle
states. We define rotational energy of a band $\kappa$ by
\begin{equation}
E_\kappa(I) = {{\left< \Phi_\kappa\right|\hat H\hat P^I_{KK} \left|
\Phi_\kappa\right>}\over {\left< \Phi_\kappa\right|\hat P^I_{KK}
\left| \Phi_\kappa\right>}} .
\end{equation}
It represents the expectation value of the Hamiltonian with respect
to a projected quasi-particle state $\kappa$. A diagram in which
rotational energies of various bands are plotted against spin $I$ is
referred to as a {\it band diagram} which contains incredibly rich
information. For example, one may observe several band-crossings at
various spins in a diagram. The first crossing is usually a crossing
between an aligning 2-qp band and the g-band, physically
corresponding to the first rotation-alignment of a high-$j$ pair of
quasiparticles.

In the present study for the transfermium region, we have found
interesting band-crossing pictures. To illustrate them, we take
representative examples, and show band diagrams in Fig. 1 for the
chain of $N=154$ isotones: $^{252}$Cf, $^{254}$Fm, and $^{256}$No.
As the neutron number is unchanged in an isotonic chain but the
proton Fermi level varies with the shell filling, the relative
position of proton and neutron Fermi levels, and therefore the SP
states in the vicinity of the Fermi levels, differ in the three. In
Fig. 1, those curves starting from 2 -- 2.5 MeV are the bands with
2-qp high-$j$ configurations (dotted and dashed curves for 2-qp
proton and 2-qp neutron configurations, respectively). At low-spins,
the g-band (0-qp configuration) is low, and is the dominant
component in the yrast wave function. However, it is seen that in
all the three diagrams, several 2-qp bands cross the g-band around
spin $I=24$ and become energetically lower after the band-crossing.
With a careful inspection, very delicate differences in the three
cases can be found: After the band-crossing, the lowest band in
$^{252}$Cf is a 2-qp band of $i_{13/2}$ protons (dotted curve) and
in $^{256}$No a 2-qp band of $j_{15/2}$ neutrons (dashed curve),
whereas in $^{254}$Fm, the lowest dotted and the dashed curves are
nearly degenerate after the band-crossing.

The above observation suggests a picture that at the band-crossing
region, proton and neutron 2-qp high-$j$ configurations compete
strongly in the yrast states at the band-crossing region. After the
band-crossing, the proton (neutron) configuration dominates the
yrast structure in $^{252}$Cf ($^{256}$No). In $^{254}$Fm, the
strongest competition between the proton and neutron configurations
is predicted. Consequences of the competition and the delicate
differences in the yrast wave functions can lead to observable
effects, particularly in the spin-dependent $g$ factor which is very
sensitive to the SP content in wave functions. In the following
three subsections, we present respectively the PSM results for the
Cf, Fm, and No isotopes.

\subsection{Cf isotopes}

\begin{figure*}
\includegraphics[width=15cm]{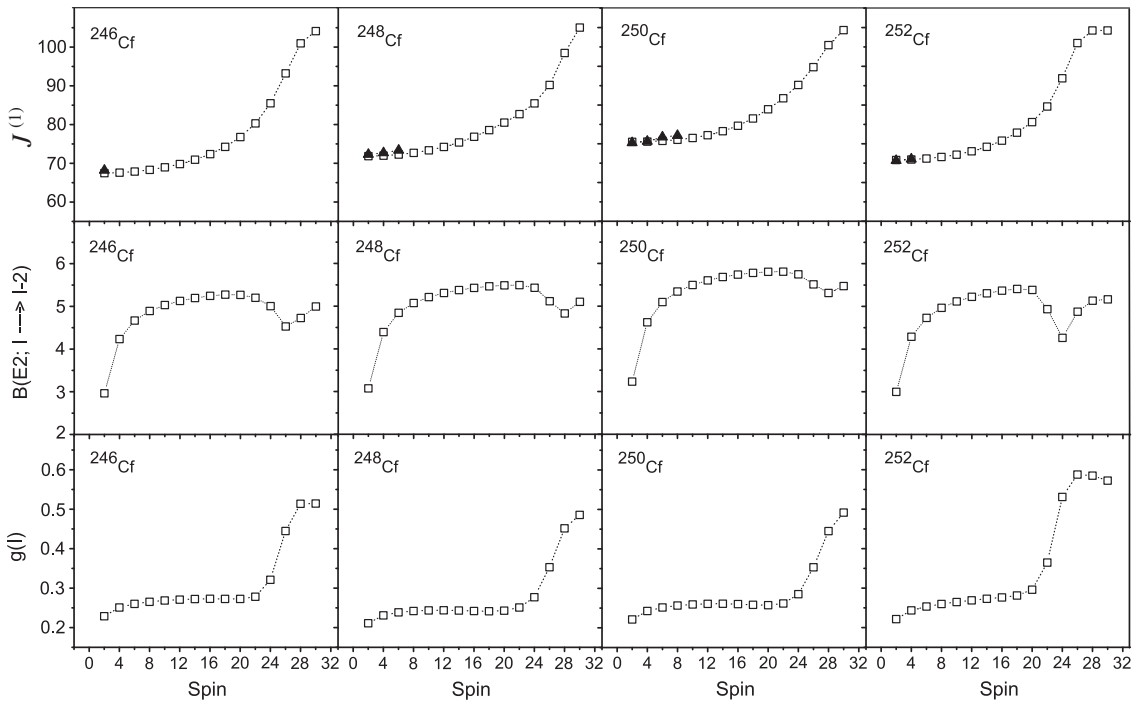}
\caption{Calculation for Californium isotopes. The top row of
figures show the calculated static moments of inertia (open
symbols), which are compared with available data (filled symbols).
The middle and bottom row are the predicted B(E2) (in $e^2b^2$) and
$g$ factor values, respectively.} \label{fig2}
\end{figure*}

We present the calculated energy levels in terms of static moment of
inertia defined by
\begin{equation}
{\cal J}^{(1)} = {{2I-1}\over {E(I)-E(I-2)}}.
\end{equation}
This quantity describes changes in band energies as spin varies.
Experimental data on rotational spectrum in the transfermium region
are still spare; we therefore compare our calculation with
experiment whatever data are available, and make predictions where
data do not exist. In addition, calculated $B(E2,I\rightarrow I-2)$
values defined in Eq. (\ref{BE2}) and $g$ factors in Eq.
(\ref{gfactor}) are also presented.

Figure 2 shows the results for four Cf isotopes. In the top row,
moments of inertia of these isotopes exhibit the following common
behavior: At the low-spin region, they increase gently with spin,
but climb more rapidly in the spin region between $I=20$ to 30,
showing a up-bending in moment of inertia. The spin region where the
up-bending occurs corresponds to the place where bands cross to each
other and rotation-alignment takes place, as discussed before. As
shown in Fig. 2, only very limited experimental data points near the
bandheads of each isotopes are known, with which our calculation
agrees well.

The cause for the rapid rise in ${\cal J}^{(1)}$ in the high-spin
region is attributed to rotation-alignment, which concerns the
nature of the crossing band(s) with aligning pairs from high-$j$
orbitals. It has already been seen in the energy-vs-spin plot (Fig.
1) that the rotationally-lowering of 2-qp bands leads to a crossing
with the g-band. Among the lowest 2-qp bands that cross the g-band
are those with 2-quasi-neutron configurations from the $j_{15/2}$
orbital and 2-quasi-proton ones from the $i_{13/2}$ orbital. These
two kinds of configurations align their spins with very competitive
probabilities. We find that for the Cf isotopes at deformation of
$\varepsilon_2 \sim 0.24$, rotation-alignment of the $i_{13/2}$
2-quasi-proton $K=1$ state (coupled by $K=5/2$ and $7/2$ $i_{13/2}$
quasi-protons) is the dominant configuration  and is responsible for
the rapid rise in moment of inertia in the Cf isotopes. This is a
theoretical interpretation, and we need further experimental
observations to support the picture.

Crossing of two bands with different configurations (here one
fully-paired configuration and one configuration with high-$j$
particle alignment) can cause a structure change for spin states
before and after the crossing, which can lead to observable effects.
Electromagnetic transition probability reflects such changes in wave
functions. In the middle and bottom rows in Fig. 2, we present
calculated B(E2) and $g$ factor values, respectively. We observe
that in all the isotopes, the B(E2) values increase smoothly with
spin (The rapid rise near the band heads is due to the geometric
Clebsch-Gordan coefficients), but with a drop around $I=26$. The
drop corresponds to smaller B(E2) values which are attributed to
different structure in wave functions of the initial and the final
state. In the $g$ factor calculation, we observe a nearly-constant
behavior for the low-spin states in all the four isotopes, but a
sudden increase in the high spin region where rotation-alignment
occurs. The rise is large, which nearly doubles the low-spin value
of $g$ factor. As only individual protons can bring positive
contribution to the $g$ factor and thus can accomplish the effect,
these results must imply a sudden increase in the proton component
in the wave function. Detailed analysis shows that for the high-spin
states with large $g$ factors, the wave functions indeed have an
increased amount of component from the proton $i_{13/2}$ orbit.
Experiment of $g$ factor at high-spins will be a strict test for our
prediction, and in turn for the Nilsson SP states employed in our
model.

\subsection{Fm isotopes}

\begin{figure*}
\includegraphics[width=15cm]{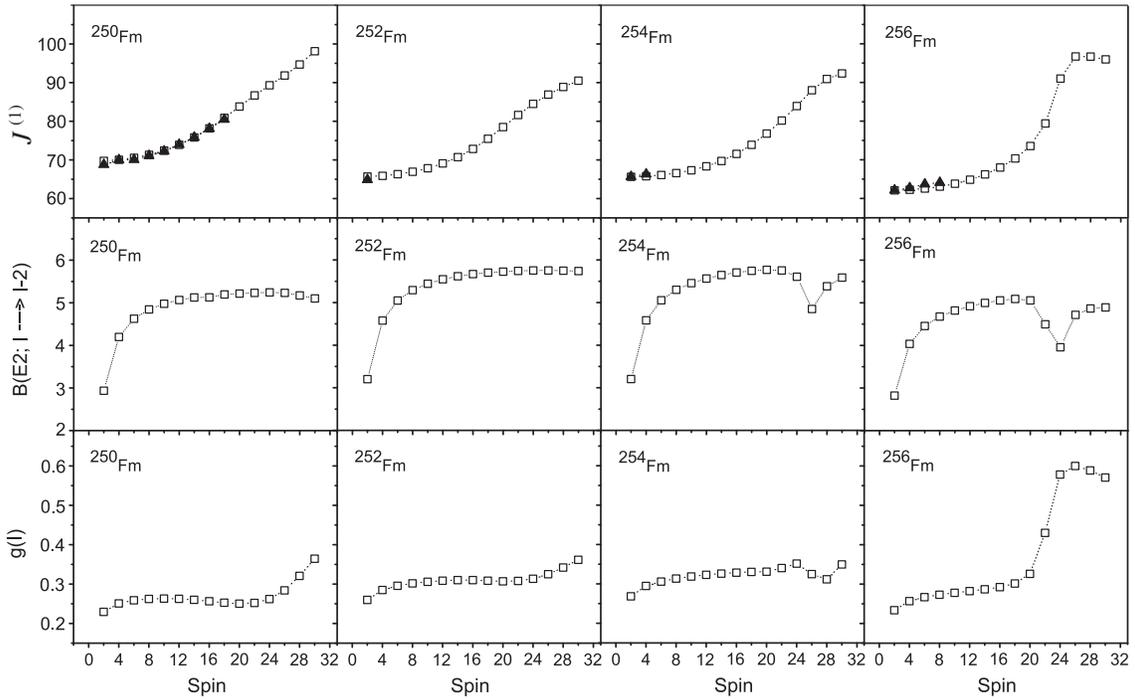}
\caption{Same as Fig.2, but for Fermium isotopes.} \label{fig3}
\end{figure*}

In Fig. 3, we show the results for four Fm isotopes. The moments of
inertia in the top row exhibit a similar behavior as in the Cf
isotopes. As can be seen, our calculation reproduces the known data
well. In particular, the yrast band in $^{250}$Fm \cite{Fm250} was
measured up to $I=18$, for which the PSM calculation yields an
excellent description. Comparing these four isotopes, the
calculation predicts a more rapid rise in ${\cal J}^{(1)}$ at
high-spins in the heavier isotope $^{256}$Fm.

It is interesting that for Fm isotopes, the lowest 2-qp bands that
cross the g-band correspond to the states with $K=1$
2-quasi-neutrons from the $j_{15/2}$ orbital (coupled by the $K=7/2$
and $9/2$ states) and $K=1$ 2-quasi-protons from the $i_{13/2}$
orbital (coupled by the $K=7/2$ and $9/2$ states), with very
competitive probability. Therefore, a pair of $j_{15/2}$ neutrons
and a pair of $i_{13/2}$ protons both contribute to
rotation-alignment, which is the cause for the irregularity in
moment of inertia in the Fm isotopes. Similar conclusion was also
obtained by the cranked relativistic mean field calculation in Ref.
\cite{Fm250}. Only in $^{256}$Fm, rotation-alignment of a pair of
$i_{13/2}$ protons is predicted by the current calculation to be
more favored over neutrons.

In the middle row of Fig. 3, a smooth behavior in B(E2) values is
obtained up to the highest spin state in the two lighter Fm
isotopes, while a drop is predicted for high spin states in the two
heavier ones. As for $g$ factors presented in the bottom row, a
common behavior is seen for all four isotopes for spin states before
the band crossing, namely, the $g$ factor values are nearly constant
with only slight variations as a function of spin. The smooth
variation continues at high spin states for $^{250,252,254}$Fm, but
for $^{256}$Fm, a sudden rise is predicted. The situation for
$^{256}$Fm is thus very similar to the Cf isotopes discussed before:
The increase in $g$ factor at high spins is contributed by the
aligning protons. Indeed, detailed analysis shows that the aligning
particles are the protons from the $i_{13/2}$ orbital coupled by the
$K=7/2$ and 9/2 states, which increases the proton component in wave
functions. On the other hand, the near-constant $g$ factors at high
spin states for $^{250,252,254}$Fm are understood as a combined
effect of proton- and neutron-alignment. The negative contribution
from neutrons to the $g$ factor is compensated by the positive
contribution from protons, leaving the total $g$ factor almost
unchanged for the entire spin region.

\subsection{No isotopes}

\begin{figure*}
\includegraphics[width=15cm]{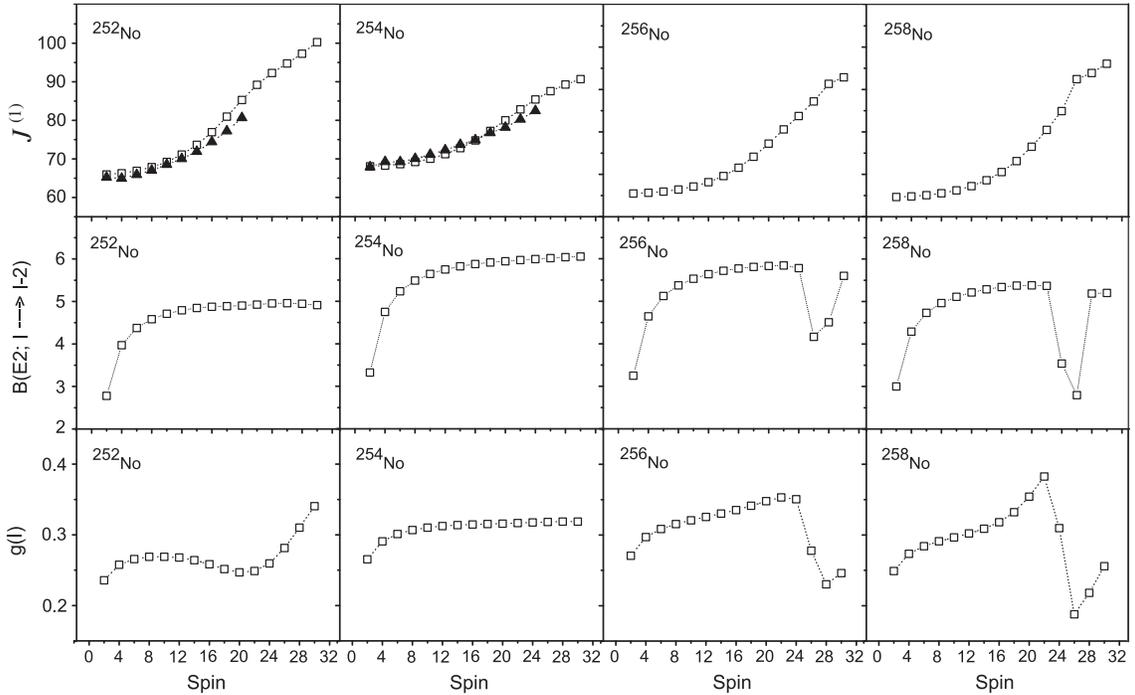}
\caption{Same as Fig. 2, but for Nobelium isotopes.} \label{fig4}
\end{figure*}

\begin{figure}
\includegraphics[width=6.0cm]{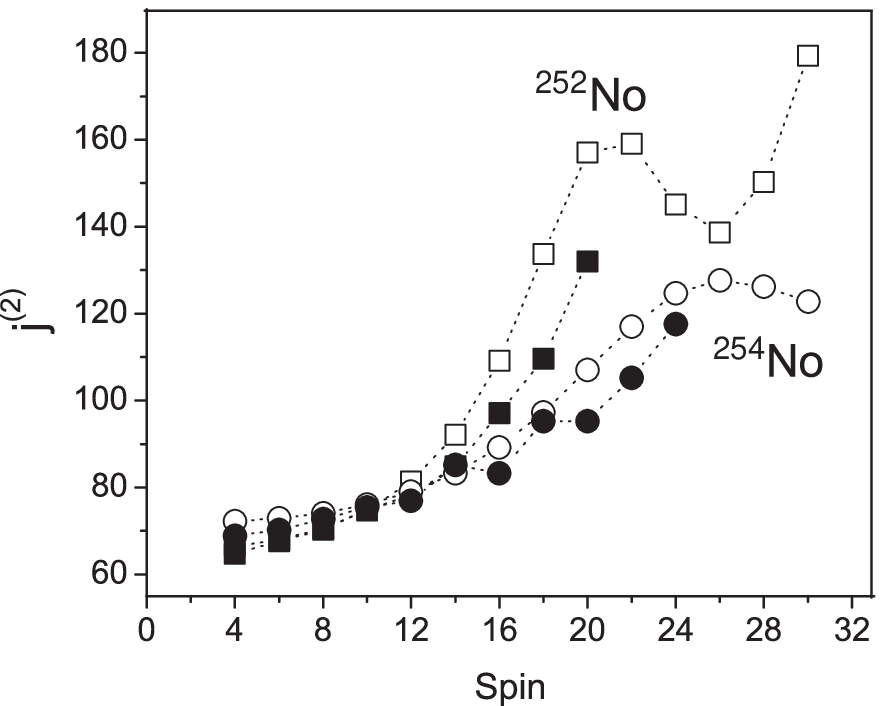}
\caption{Comparison of calculated dynamic moments of inertia ${\cal
J}^{(2)}$ (open symbols) with experimental ones (filled symbols) for
$^{252,254}$No. ${\cal J}^{(2)}(I)$ is defined as $4/\Delta
E_\gamma(I)$, with $\Delta
E_\gamma(I)=E_\gamma(I)-E_\gamma(I-2)=E(I)-2E(I-2)-E(I-4)$.}
\label{fig5}
\end{figure}

We consider four No isotopes $^{252-258}$No. Yrast bands of the two
lighter No isotopes $^{252,254}$No were measured up to considerably
high spins \cite{No252,No254}. These two nuclei are also the most
theoretically studied examples. Egido and Robledo \cite{ER00} found
in their cranked Hartree-Fock-Bogoliubov calculation with the Gogny
force that the first upbending in moment of inertia in $^{254}$No is
attributed to alignment of a pair of $i_{13/2}$ protons. The cranked
relativistic Hartree-Bogoliubov calculation \cite{Afa03} suggested
that a simultaneous alignment of the $i_{13/2}$ proton and
$j_{15/2}$ neutron pairs is responsible for the changes in moments
of inertia in both $^{252,254}$No.

Fig. 4 shows the PSM results for four No isotopes. In the top row of
Fig. 4, a comparison with the known $^{252,254}$No data in terms of
static moments of inertia ${\cal J}^{(1)}$ yields a good agreement.
Beyond the existing data points, a continued steady increase in
${\cal J}^{(1)}$ is predicted by the present calculation for high
spin states, which does not seemingly indicate an anomaly in moment
of inertia caused by band-crossing. A more sensitive plot for
dynamic moments of inertia ${\cal J}^{(2)}$ for these tow isotopes
is shown in Fig. 5, in which the calculation suggests a down-turning
in slope of ${\cal J}^{(2)}$ at $I=24$ for $^{254}$No. In
$^{252}$No, the calculated ${\cal J}^{(2)}$ shows a down-turning at
$I=22$, followed by a up-turning at $I=26$. The current data points
for both nuclei stop just before the turning points. Extension of
experimental measurement to higher spins will test our prediction.

In the middle row of Fig. 4, the calculation predicts a smooth trend
in B(E2) for the two lighter isotopes $^{252,254}$No. Similar B(E2)
behavior is obtained for their isotones $^{250,252}$Fm (see Fig. 3).
For the two heavier isotopes $^{256,258}$No, a sharp drop in B(E2)
is predicted at $I=26$. Drop in B(E2) corresponds to a structure
change in the yrast band caused by rotation-alignment; however the
B(E2) values alone cannot distinguish whether the cause is due to
protons or neutrons. In the bottom row of Fig. 4, we observe the
following evolution in $g$ factor as neutron number varies. In
$^{252}$No, we predict an increase in $g$ factor starting from
$I=22$. As discussed before, the $i_{13/2}$ proton alignment is
responsible for this behavior. In the next isotope $^{254}$No,
rather constant $g$ factor is predicted, which is understood as in
the case of $^{250,252,254}$Fm where a canceling in the proton- and
neutron-contribution to the $g$ factor is expected. Moving to the
heavier No isotopes, we predict a big drop in $g$ factor at about
$I=24$. The reason for the decrease in $g$ factor is due to a large
component of the 2-quasi-neutron $K=1$ state (coupled by the $K=7/2$
and $9/2$ $j_{15/2}$ states) in the wave functions at and after the
band-crossing. This suggests that for $^{256,258}$No, the $j_{15/2}$
neutrons win in the proton-neutron alignment competition, and the
high-$j$ neutron component is dominant in the yrast wave functions
at high spin states.

\section{Discussions}

The early work of Egido and Ring \cite{ER84} studied the actinides
nuclei using the rotating shell model with particle number
projection and inclusion of quadrupole-pairing interaction. For the
heaviest isotopes they studied, they obtained similar predictions
for $^{244,246}$Cf (in particular for $g$ factors) as ours. They
reached the same conclusion that there is a competition between
proton and neutron alignment. For the alignment low $K$ levels in
the high $j$ orbits are important. Filling in more and more
particles one therefore has to provide more and more energy in order
to get alignment. In this mass region the neutron $j_{15/2}$ orbital
contains generally more particles than the proton $i_{13/2}$
orbital. Therefore the alignment of protons is generally favored as
compared to alignment of neutrons. That is what we have observed in
all the four Cf isotopes as well as in $^{256}$Fm and $^{252}$No.
Since nuclei are complex many-body systems, situation near the Fermi
levels can change with neutron and proton numbers and a reversed
case may occur. Moving away from heavier elements than what were
discussed by Egido and Ring \cite{ER84}, we have found examples in
which neutron alignment is more favored. In the heavier No isotopes,
the present calculation suggests such examples.

However, from the energy levels alone we cannot easily distinguish
proton and neutron contributions and a gradual, smooth alignment can
hardly disentangled from the variable moment of inertia in the
reference. B(E2) values cannot distinguish proton and neutron
contributions either because one cannot experimentally separate the
proton and neutron contributions to B(E2). A better probe in this
regard is the magnetic moments through the study of $g$ factors. $g$
factors show rather clearly which kind of particle is aligning even
in cases where no clear changes can be seen in the sequence of
energy levels and in B(E2).

Our results presented in Figs. 2, 3, and 4, in particular those of
$g$ factor calculation, suggest a competition picture in
rotation-alignment between the high-$j$ intruder neutrons and
protons. The delicate changes in wave functions result in distinct
observables which can be tested in future experiment. These changes
reflect nature of the rotation-aligning configurations and the
characters of SP states for the transfermium region.

We have thus found measurable quantities that are sensitive to SP
states, and therefore can serve as a testing ground. It is important
to comment on the Nilsson SP states in the present calculation.
Strictly speaking, SP states in deformed nuclei are not directly
measurable. They are produced by calculations, and are
model-dependent. Single-particle states in our shell-model basis are
constructed by using the deformed Nilsson potential. Therefore, the
above results and discussions depend on the Nilsson SP states. The
adopted Nilsson parameters are the 1985 parameterisation of
Bengtsson and Ragnarsson \cite{BR85}. For the mass region of the
present interest, this parameterisation gives a SP distribution very
similar to another popular set of SP states produced by the
Woods-Saxon potential of Chasman {\it et al.} \cite{Chas77}; the
latter was used to assign configurations of the observed level
structure in odd-mass transfermium nuclei
\cite{Odd-mass1,Odd-mass2}. Thus in a sense, experimental
confirmation or repudiation of the present PSM results is a test of
the standard Nilsson model for the transfermium mass region.

\section{Summary}

One of the key ingredients for locating the superheavy "island of
stability" is the description of SP states. To predict SP states for
SHE, extrapolation from existing microscopical models for the stable
mass region is often assumed. However, the so-obtained SP states
need to be carefully checked with experiment. Low-spin ground-state
band in such heavy, deformed systems exhibits nothing but a
collective rotor behavior, thus telling little useful information on
SP states. To extract useful information from observables, one needs
to study those excited configurations that directly carry
information on individual particles. The present work attempts to
understand the SP structure through the study of measurable
quantities related to 2-qp excitations with rotation-alignment. The
study has used the Projected Shell Model, which adopted the
widely-used deformed Nilsson SP states as a starting basis,
systematically performed shell model calculations for some Cf, Fm,
and No isotopes, and compared the results with existing data in
terms of moment of inertia. The calculation has further predicted
B(E2) and $g$ factor values for these nuclei.

Results of the present calculation can be summarized as follows. The
static moments of inertia of all isotopes studied in this paper has
been shown to have a similar behavior: At low-spin region, they
increase gently with rotation, but a more rapid rise is seen in the
spin region $I=20$ to 30. The rise of ${\cal J}^{(1)}$ indicates
additional contribution of angular momentum from the aligning
quasiparticles. By studying band diagrams, we have found that the
relevant quasiparticles are those from the high-$j$ intruder
orbitals with $i_{13/2}$ for protons and $j_{15/2}$ for neutrons.
For all the three nuclei ($^{250}$Fm, $^{252}$No, and $^{254}$No)
for which there exist longer sequence of experimental levels, our
calculation has obtained a good agreement with data. We have found
that the experimental states were measured just before the
band-crossing spin. The prediction for the higher spin states awaits
future experimental confirmation.

Electromagnetic transition properties reflect microscopic inside of
wave functions, and thus are more sensitive probes for the
structure. Study of energy levels alone cannot fully understand the
microscopic origin of the rotation-alignment, and the variation in
moment of inertia and B(E2). By studying $g$ factors, we have found
important differences in aligning particles. The proton-$i_{13/2}$
alignment is preferred for the Cf isotopes, as well as for
$^{256}$Fm. The strongest competition in rotation-alignment between
the high-$j$ protons and neutrons has been suggested for the lighter
Fm and No isotopes, in which $g$ factors are predicted to stay
nearly constant in the spin region where rotation-alignment takes
place. Finally, the neutron-$j_{15/2}$ rotation-alignment is favored
in the heavier No isotopes. It is expected that the future
experiment, in particular $g$ factor measurements, will provide a
strict test for the current theory, and for the applicability of the
standard Nilsson scheme \cite{BR85} (also of the Woods-Saxon model
\cite{Chas77} as its results are very similar to those of the
Nilsson model) to the superheavy mass region.

Y.S. thanks the colleagues at Institute of Modern Physics, Tsinghua
University, and Peking University, China, for the warm hospitality
extended to him. This work was supported in part by the National
Natural Science Foundation of China under contract No. 10325521 and
10875077, and by the Chinese Major State Basic Research Development
Program through grant 2007CB815005.

\baselineskip = 14pt
\bibliographystyle{unsrt}

\end{document}